\documentclass[acmsmall,screen]{acmart}

\usepackage{tikz}
\usetikzlibrary{positioning}
\usepackage{csquotes}
\usepackage{xcolor}
\usepackage{fontawesome5}
\usepackage{needspace}

\setcopyright{none}
\renewcommand\footnotetextcopyrightpermission[1]{}
\copyrightyear{2026}
\acmYear{2026}
\acmDOI{}
\acmVolume{}
\acmNumber{}
\acmArticle{}

\newcommand{\publicationstatus}{Accepted for publication in Communications of the ACM}
\makeatletter
\def\@journalName{Communications of the ACM}%
\def\@journalNameShort{Commun. ACM}%
\def\@acmPubDate{}%
\renewcommand{\@mkbibcitation}{%
  \begingroup
  \let\@vspace\@vspace@orig
  \let\@vspacer\@vspacer@orig
  \par\medskip\small\noindent{\bfseries ACM Reference Format:}\par\nobreak
  \noindent\begingroup\def\\{\unskip{}, \ignorespaces}\authors\endgroup.
  \@title. Accepted for publication in \textit{\@journalName}.
  \par\endgroup
}
\makeatother

\AtBeginDocument{%
  \fancypagestyle{standardpagestyle}{%
    \fancyhf{}%
    \fancyhead[LE,RO]{\footnotesize\thepage}%
    \fancyhead[LO]{\footnotesize\shorttitle}%
    \fancyhead[RE]{\footnotesize\shortauthors}%
    \fancyfoot[C]{\footnotesize\publicationstatus}%
  }%
  \fancypagestyle{firstpagestyle}{%
    \fancyhf{}%
    \fancyhead[L]{\footnotesize Accepted manuscript}%
    \fancyhead[R]{\footnotesize\thepage}%
    \fancyfoot[C]{\footnotesize\publicationstatus}%
  }%
  \pagestyle{standardpagestyle}%
}

\begin{document}

\title{AI Slop and the Software Commons}

\author{Sebastian Baltes}
\email{sebastian.baltes@uni-heidelberg.de}
\affiliation{%
  \institution{Heidelberg University}
  \city{Heidelberg}
  \country{Germany}
}

\author{Marc Cheong}
\email{marc.cheong@unimelb.edu.au}
\affiliation{%
  \institution{University of Melbourne}
  \city{Melbourne}
  \country{Australia}
}

\author{Christoph Treude}
\email{ctreude@smu.edu.sg}
\affiliation{%
  \institution{Singapore Management University}
  \city{Singapore}
  \country{Singapore}
}

\renewcommand{\shortauthors}{Baltes et al.}

\begin{abstract}
We argue that \textit{AI slop} in software development is creating a tragedy of the commons. Slop is cheap to generate but expensive to review, and its costs fall on shared resources such as reviewer capacity, which was limited even before AI assistants became widespread. Ostrom documented how communities avoid such tragedies by setting and enforcing their own rules. We translate Ostrom's design principles for sustainable commons institutions into concrete proposals for contributors, reviewers, tool developers, team leads, organizational leaders, and educators.
\end{abstract}




\maketitle


In January 2026, the \textit{curl} project shut down its HackerOne bug-bounty program because a flood of plausible but unfounded AI-generated vulnerability reports consumed too much reviewer time.\footnote{Daniel Stenberg, \href{https://daniel.haxx.se/blog/2026/01/26/the-end-of-the-curl-bug-bounty/}{\textit{The End of the curl Bug-Bounty}}, January 26, 2026.} Such reports are a form of \textit{AI slop}.
\citeauthor{kommers2026slop}~\cite{kommers2026slop} identified three prototypical properties that slop shares across domains: \emph{``superficial competence''} (a veneer of quality without substance), \emph{``asymmetric effort''} (creation requires little of the effort needed without AI), and \emph{``mass producibility.''}

Our position is that \textit{AI slop} in software development constitutes a tragedy of the commons~\cite{hardin1968tragedy}.
In Hardin's example, each herder benefits from adding cattle to a shared pasture while distributing the costs of overgrazing across everyone who uses it, and this individually rational behavior leads to collective ruin.
AI-assisted contributions can produce the same imbalance: Contributors gain from rapid generation while drawing on the finite time and expertise that others provide to check, correct, and maintain the output. In online discussions of \textit{AI slop}, developers described risks to reviewer capacity, codebase integrity, knowledge resources, collaborative trust, and the talent pipeline~\cite{baltes2026slop}.

\section{Evidence from the Discourse}
\label{sec:evidence}

In a previous empirical study~\cite{baltes2026slop}, we analyzed 1{,}154 posts across 15 discussion threads from Reddit and Hacker News in which developers explicitly used the phrase \textit{AI slop}.

\textbf{Reviewers bear the cost of slop.}
A contributor can generate slop cheaply, but checking it takes up the time of a reviewer with the necessary expertise. Economists call such a cost an \textit{externality}, because it falls on someone other than the person who caused it. One team reported 30 pull requests per day split across six reviewers.
One reviewer wrote: \emph{``They're literally just using you to do their job, i.e., critically evaluate and understand their \textit{AI slop} and give it the next prompt.''}
\label{para:curl-update}Even AI-assisted reports that identify real problems can take up too much reviewer time. In October 2026, \textit{curl} maintainer Daniel Stenberg reported that hallucinated security reports had largely disappeared, but accurate AI-assisted reports still arrived in high volume and were often hundreds of lines long, convoluted, and repetitive.\footnote{Daniel Stenberg, \href{https://daniel.haxx.se/blog/2026/10/22/ignore-the-slop/}{\textit{Ignore the Slop}}, October 22, 2026.}
Stenberg does not consider these reports slop because they identify real problems, but in our view their verbosity is slop because it adds review effort without helping reviewers assess the finding. On the Linux kernel security list, different people ran the same AI tools and reported the same bugs, which made the list \emph{``almost entirely unmanageable,''} according to Linus Torvalds.\footnote{Linus Torvalds, \href{https://lkml.org/lkml/2026/5/17/896}{\textit{Linux 7.1-rc4}}, LKML, May 17, 2026.}

\textbf{Slop degrades codebases and knowledge resources.}
Merging AI output without careful review can add technical debt. A model might mask a race condition with a \texttt{setTimeout} delay, silence a type mismatch with a cast to \texttt{any}, or rewrite a test to accept broken behavior. In one case, a developer reported \emph{``AI hallucinating external services, then mocking out the hallucinated external services,''} producing an internally coherent but fictional integration.
Beyond the codebase, developers reported that public tutorials and documentation were degrading. Learners who rely on these resources risk copying errors before they can recognize them. Model vendors use these resources as training data, and \citeauthor{shumailov2024model} found that repeatedly training models on outputs from earlier models can cause irreversible defects in the resulting models, an effect that they called \textit{model collapse}~\cite{shumailov2024model}.
Vendors of tools that produce slop therefore risk their own tragedy of the commons if their output degrades the shared data used to train future models.

\textbf{Incentives reward slop.}
Developers described incentives that reward quantity over quality, such as contribution graphs and search-engine rankings. Besides, bug bounties create a monetary incentive to submit as many AI-generated reports as possible.
One Reddit commenter posted a prediction about GitHub's AI features: \emph{``This will just be weaponized by people trying to boost their profile in an ironically shrinking job market.''}
Developers also described corporate mandates to adopt AI tools regardless of fit, including C-level executives \emph{``running parts of our codebase through AI tools and literally copy pasting the response as an answer to every technical problem.''}
Like Hardin's herders~\cite{hardin1968tragedy}, contributors benefit from each additional contribution, and the incentives mentioned above increase the pressure to submit more contributions for reviewers to check.

\Needspace{8\baselineskip}
\section{The Commons Under Strain}
\label{sec:roles}

Reviewer capacity, defined as the time and expertise available to assess contributions, is a finite shared resource (\autoref{fig:commons}). The same developer can submit a contribution and review someone else's. A contribution that the contributor has verified can improve codebases and knowledge resources, whereas an unverified contribution leaves the verification to a reviewer. Contributors should therefore verify their changes before requesting a review, or they risk losing reviewers' trust and willingness to review their work.

\begin{figure}[!t]
  \centering
  \begin{tikzpicture}[
  font=\scriptsize,
  box/.style={draw, rounded corners, align=center,
              inner xsep=4pt, inner ysep=4pt,
              execute at begin node={\hyphenpenalty=10000\exhyphenpenalty=10000}},
  main/.style={box, fill=gray!5, minimum height=1.8cm},
  actor/.style={box, dashed, minimum height=0.6cm},
  edge/.style={-latex, thick},
]
  \node[main, text width=3.0cm] (P) at (0,0) {\textbf{Contributors}\\[2pt]developers who submit code, bug reports, or documentation\\[2pt](\textit{Hardin's herders})};
  \node[main, text width=4.0cm] (C) at (8.2,0) {\textbf{Shared resources}\\[2pt]reviewer capacity, codebase integrity, knowledge resources, collaborative trust, talent pipeline\\[2pt](\textit{Hardin's pasture})};
  \coordinate (top) at (0,0.55);
  \coordinate (mid) at (0,0);
  \coordinate (low) at (0,-0.4);
  \draw[edge] (C.west |- top) -- node[midway,above,align=center]{private gains:\\faster output, contribution counts} (P.east |- top);
  \draw[edge] (P.east |- mid) -- node[midway,above]{slop consumes reviewer capacity} (C.west |- mid);
  \draw[edge] (P.east |- low) -- node[midway,below,align=center]{verified contributions improve\\codebases and knowledge resources} (C.west |- low);
  \node[actor] (E) at (-1.1,-2.15) {\textbf{Educators}};
  \node[actor] (T) at (1.0,-2.15) {\textbf{Tool developers}};
  \node[actor] (L) at (6.0,-2.15) {\textbf{Team leads and organizational leaders}};
  \coordinate (lp) at (4.2,0);
  \coordinate (pl) at (1.45,0);
  \coordinate (lc) at (7.8,0);
  \draw[edge,dashed] (E.north) -- node[midway,left,align=right]{teach\\verification} (E.north |- P.south);
  \draw[edge,dashed] (T.north) -- node[midway,left,align=right]{make output\\inspectable} (T.north |- P.south);
  \draw[edge,dashed,rounded corners=2pt] (L.north -| lp) -- (4.2,-1.62) -- node[midway,above]{stop rewarding volume} (1.45,-1.62) -- (P.south -| pl);
  \draw[edge,dashed] (L.north -| lc) -- node[midway,right,align=left]{reserve time\\for review} (C.south -| lc);
\end{tikzpicture}
  \Description{A Contributors box on the left, labeled as developers who submit code, bug reports, or documentation and as Hardin's herders, and a Shared resources box on the right, listing reviewer capacity, codebase integrity, knowledge resources, collaborative trust, and the talent pipeline and labeled as Hardin's pasture. A solid arrow from the shared resources to the contributors is labeled private gains: faster output, contribution counts. A solid arrow from the contributors to the shared resources states that slop consumes reviewer capacity. A second solid arrow from the contributors to the shared resources states that verified contributions improve codebases and knowledge resources. Below, dashed boxes for educators, tool developers, and team leads and organizational leaders have dashed arrows. Educators teach verification and tool developers make output inspectable, both pointing to the contributors. Team leads and organizational leaders stop rewarding volume, pointing to the contributors, and reserve time for review, pointing to the shared resources.}
  \caption{Contributors and shared resources in the software commons, with their counterparts in Hardin's example~\cite{hardin1968tragedy}. Reviewer capacity represents the time and expertise of reviewers; the same developer can be both contributor and reviewer. Dashed boxes and arrows show the actions that we propose for other stakeholders.}
  \label{fig:commons}
\end{figure}

In software development, asymmetric effort and mass producibility~\cite{kommers2026slop} let contributors generate more changes than reviewers have time to check. Under the resulting time pressure, reviewers can miss defects, because the superficial competence of slop makes generated code look correct.
For example, a developer who submits an AI-generated pull request without checking it produces it faster and raises their contribution count. A reviewer who notices a test rewritten to accept broken behavior, a maintainer who later diagnoses a regression, and a junior developer who copies a broken tutorial all pay for it.
In open-source projects, these reviewers and maintainers are often volunteers with limited time. \citeauthor{eghbal2020working} documented that maintainers were under-resourced before AI assistants became widespread~\cite{eghbal2020working}.
In industry, the mandates and volume-based rewards described in Section~\ref{sec:evidence} increase the number of pull requests without increasing the number of reviewers~\cite{baltes2026slop}.

Reviewer capacity and other shared resources also depend on decisions made outside contribution and review. For example, \textit{tool developers} build the generation tools with which contributors produce changes, such as coding agents and editor integrations. They also build the repository tools with which reviewers assess these changes, such as code review and continuous integration (CI) systems. \textit{Team leads and organizational leaders} assign tasks and decide how developers' work is evaluated, and \textit{educators} teach students and developers how to write and check code. The dashed arrows in \autoref{fig:commons} show the actions that we propose for these groups, and Section~\ref{sec:governance} describes these actions in more detail.

\Needspace{8\baselineskip}
\section{Preventing the Collapse}
\label{sec:governance}

\citeauthor{hardin1968tragedy}'s example showed how users can overuse a shared resource, but not how communities can prevent the overuse. We draw on \citeauthor{ostrom1990governing}'s empirical work on commons governance~\cite{ostrom1990governing}, which documents communities that sustained shared resources through rules they set themselves rather than relying solely on privatization or top-down control.
We translate Ostrom's eight design principles into proposals to prevent the overuse of shared resources in software projects, using adapted names of the principles as paragraph headings. Our running example is a contributor who prepares an AI-assisted bug fix. Before requesting a review, the contributor checks the diff manually, explains the intended behavior, and provides a regression test.

\textbf{Clearly defined boundaries.}
\label{para:provenance}
Projects should define evidence requirements, that is, what a contribution must include before a reviewer assesses it. For pull requests, we propose a form that asks contributors (1) which AI tools they used, (2) which parts of the contribution they checked manually, and (3) how they verified and validated the contribution, for example with tests and static analysis. It is enough for contributors to name the AI tools that they used, because review depends on whether they verified and validated the contribution, not on which tool wrote which part of it. For bug reports, maintainers should require a test case or proof of concept that reproduces the bug, and they can ask contributors to describe the bug briefly in their own words.
Generation tools already keep session logs of prompts and generated changes, and tool developers should link these logs to pull requests automatically, so that reviewers can consult them when a change needs explanation. The Entire CLI, for example, pushes each agent session log to the repository and links it to the resulting commit.\footnote{Entire, \href{https://github.com/entireio/cli}{\textit{Entire CLI}}, accessed October 31, 2026.}
Session logs miss AI use in tools that do not record sessions, such as code pasted from a chat assistant in a browser, and nobody can verify that contributors disclosed all AI use. The evidence requirements should therefore apply to every contribution, regardless of declared origin. Because walkthroughs take time from both the contributor and the reviewer, maintainers should normally reserve them for first-time contributors and contributors whose earlier contributions contained errors or unexplained changes. Reviewers should also be able to request a walkthrough of any contribution when a specific concern remains. In a walkthrough, the contributor would explain the changed lines, the intended behavior, and how the tests demonstrate it.

\textbf{Congruence between rules and local conditions.}
\label{para:local-rules}
Once a project has defined its evidence requirements, team leads should adapt them depending on the risk of a change. A documentation correction may need only a short description, whereas a change to authentication code needs tests and a detailed explanation. When reviewer capacity is low, they should require tests and a detailed explanation for all code changes, not only for high-risk ones. Team leads should also measure review effort, rework, and post-merge defects for the project as a whole at regular intervals, using data such as review times, the number of review rounds, and bug reports linked to merged pull requests. The results show whether the evidence requirements catch defects without overloading reviewers. Using them to evaluate individual developers would be counterproductive, because once a measure determines rewards, developers work toward the measure instead of the goal behind it, as Goodhart's law describes. Rewards for more pull requests or more AI use, for example, do not lead to better contributions. Such rewards also credit contributors for their pull requests but not reviewers for checking these pull requests. Organizational leaders should therefore reserve time for review when assigning work and credit reviewing in performance evaluations.

\textbf{Collective-choice arrangements.}
\label{para:collective-choice}
Contributors and reviewers should together decide on the evidence requirements introduced above and record them in their contributing guide or team agreement. They can start from norms that developers already follow, such as the norm that one developer stated as \emph{``It's not AI's code, it's my code.''}
Recorded as a rule, the norm makes the contributor of the AI-assisted bug fix in our running example responsible for checking and explaining the generated changes. When the evidence requirements miss defects or overload reviewers, contributors and reviewers should both be able to propose changes to the requirements, for example by opening an issue in an open-source project or by raising the topic in a team meeting.

\textbf{Monitoring.}
\label{para:monitoring}
To help reviewers monitor what enters a codebase, repository tools should show them the evidence outlined above next to the diff. When a pull request changes high-risk code, repository tools should request a review from the developers responsible for that code. GitHub already offers pull-request templates, \texttt{CODEOWNERS} files that name required reviewers for specific files or directories, and branch protection rules that require passing checks and approvals before merging.\footnote{GitHub, \href{https://docs.github.com/en/pull-requests/reference/managing-and-standardizing-pull-requests}{\textit{Managing and standardizing pull requests}}, accessed October 22, 2026.}
Generated tests can show the same superficial competence as generated code, passing and looking like verification while sharing the mistaken assumptions of the code under test~\cite{zhao2026misguidance}.
Repository tools should therefore also check whether the submitted tests can catch errors, for example by running a regression test for a bug fix on the code before the fix, where it should fail, or by mutation testing. Reviewers can then spend their time on problems that tools cannot detect.

\label{para:ai-review}
Beyond automated checks, AI review tools could take over part of the monitoring that reviewers do. GitHub's Copilot code review posts comments on pull requests and proposes fixes, but GitHub warns that it can produce false positives and false negatives.\footnote{GitHub, \href{https://docs.github.com/en/copilot/responsible-use/agents}{\textit{Application card: GitHub Copilot Agents}}, accessed October 22, 2026.}
Reviewers therefore have to verify every comment, and the comments add work if verifying them takes longer than reviewing the change without the tool~\cite{cihan2025automated}.
A tool that reviews AI-generated code may also repeat the mistakes of the tool that generated it. We therefore propose AI assistance for tasks with results that reviewers can check quickly, such as summarizing changes, grouping duplicate reports, and checking a contributor's claims against tests or static analysis, with each result linked to its evidence. Projects should judge AI assistance by the measurements proposed above, adding the time spent generating code and the useful contributions that the assistance wrongly rejects. Maintainers should let a tool act without human review, for example by closing duplicate reports, only after measurements in their project show that its error rate is acceptable.
\label{para:changing-models}Better models may reduce the effort of generating and reviewing each contribution, but as long as a human has to check or approve any part of a contribution, the total review time grows with the number of contributions, whereas generation is cheap.

\textbf{Graduated sanctions.}
\label{para:sanctions}
Maintainers should return incomplete contributions to the contributor with a request to add the missing explanation or evidence. If a contributor repeatedly violates the project's contribution rules, for example by deliberately misrepresenting AI use, maintainers should limit the contributor's open pull requests or revoke their merge rights. The project's contributing guide should list the problems for which maintainers return a contribution and the violations for which they limit or revoke a contributor's rights. The same rules should apply to contributions with and without AI assistance.

\textbf{Conflict-resolution mechanisms.}
\label{para:walkthrough}
A reviewer who cannot understand part of a contribution should be able to request a walkthrough of that part before deciding whether to approve the contribution. The project's contributing guide should state what reviewers can question in this way and how quickly and in what form contributors have to respond. If the walkthrough does not address the reviewer's concerns, they can reject the contribution. Contributors who consider a walkthrough request or a rejection unjustified should be able to ask a designated maintainer or team lead to decide.

\textbf{Recognized rights to organize.}
\label{para:organize}
Projects have already changed their contribution rules in response to slop, as \textit{curl} did by ending its bug bounty program. Such rules work only if the project's maintainers enforce them and the platforms and companies around the project support them. Code hosting platforms should give maintainers control over incoming contributions, as GitHub began to do in 2026 in response to growing numbers of low-quality pull requests.\footnote{GitHub, \href{https://github.blog/changelog/2026-06-17-limit-open-pull-requests-for-users-without-write-access/}{\textit{Limit open pull requests for users without write access}}, June 17, 2026.}
Companies that depend on an open-source project should require their developers to follow the project's contribution rules. Within companies, organizational leaders should let teams decide how they use AI tools instead of mandating their adoption.

Software engineering educators should likewise decide in which assignments and exams students may use AI tools, because AI use can impair learning~\cite{shen2026skill} and AI-generated code can contain security vulnerabilities~\cite{fu2025security}.
Students therefore need to learn to find functional defects and security vulnerabilities in code that they did not write, which educators can test in oral examinations or live coding without AI tools.
After graduation, junior developers need further practice, and organizational leaders should give them working time to review contributions together with experienced reviewers. The talent pipeline, that is, the future supply of experienced reviewers, depends on whether junior developers learn this skill.

\textbf{Nested governance.}
\label{para:nested}
Some causes of slop lie outside any single project, such as coding agents that open large pull requests without explanations, rewards for pull-request counts, and courses in which students never learn to check code that they did not write. Projects therefore need to communicate their problems, with the measurements proposed above as evidence, to the tool developers, organizational leaders, and educators who are responsible for these causes. Software foundations could collect the measurements across projects, but many depend on donations and volunteers, and companies and public funders that rely on open-source software should consider funding them for this task.

\section*{Conclusion}

AI slop creates a tragedy of the commons in Hardin's sense~\cite{hardin1968tragedy}, because contributors benefit from cheap generation while reviewers bear the cost of checking the output. Ostrom showed that communities avoid this tragedy when the people who use a shared resource set and enforce their own rules~\cite{ostrom1990governing}. In our previous study, developers described projects that had begun to do so, for example by limiting pull-request size, refusing merges, or requesting explanations of code. Projects that adopt our proposals should measure review effort, rework, and post-merge defects before and after adoption, so that researchers can compare the effects across projects. To enforce their contribution rules, projects need support from tool developers, team leads, organizational leaders, and educators, who should make verification part of their tools, workloads, and curricula. Organizations should also let their developers spend working time on reviews in the open-source projects that they depend on. Without that support, the rising number of contributions risks exhausting reviewer capacity and bringing projects closer to Hardin's proverbial \emph{``ruin.''}

\subsection*{Acknowledgments}
\label{sec:acknowledgments}
We used Claude Code and OpenAI Codex to help draft and revise the text and prepare the figure. The authors are responsible for the final content.
\label{para:acknowledgments-end}

\bibliographystyle{ACM-Reference-Format}
\bibliography{literature}

\end{document}